\documentclass[12pt]{article}

\usepackage{graphicx}
\usepackage{float}
\usepackage{cite}
\usepackage{amsfonts}
\usepackage{amssymb}
\usepackage{relsize}
\usepackage[compatibility=false]{caption}
\usepackage{subcaption}
\usepackage{xcolor}

\def\Fint{\rlap{$\Biggl\rfloor$}\Biggl\lceil}
\def\gtwid{\mathrel{\raise.3ex\hbox{$>$\kern-.75em\lower1ex\hbox{$\sim$}}}}
\def\ltwid{\mathrel{\raise.3ex\hbox{$<$\kern-.75em\lower1ex\hbox{$\sim$}}}}
\def\square{\kern1pt\vbox{\hrule height 1.2pt\hbox{\vrule width 1.2pt\hskip 3pt
   \vbox{\vskip 6pt}\hskip 3pt\vrule width 0.6pt}\hrule height 0.6pt}\kern1pt}

\begin{document}

\begin{titlepage}

\begin{flushright}
UFIFT-QG-23-9
\end{flushright}

\vskip 2cm

\begin{center}
{\bf The Need to Renormalize the Cosmological Constant}
\end{center}

\vskip 1cm

\begin{center}
N. C. Tsamis$^{1*}$, R. P. Woodard$^{2\dagger}$ and B. Yesilyurt$^{2\ddagger}$
\end{center}

\begin{center}
\it{$^{1}$ Institute of Theoretical \& Computational Physics \\
Department of Physics, University of Crete \\
GR-710 03 Heraklion, HELLAS}
\end{center}

\begin{center}
\it{$^{2}$ Department of Physics, University of Florida \\
Gainesville, FL 32611, UNITED STATES}
\end{center}

\vspace{1cm}

\begin{center}
ABSTRACT
\end{center}
We consider the massless, minimally coupled scalar on de Sitter background.
Although the 1-loop divergences of the graviton 1PI 2-point function are 
canceled by the usual Weyl ($C^2$) and Eddington ($R^2$) counterterms, 
there is still a finite, nonzero contribution to the graviton 1-point 
function. Unless this is canceled by a finite renormalization of the 
cosmological constant, the 1PI 2-point function will not be conserved, nor 
will the parameter ``$H$'' correspond to the actual Hubble constant. We 
argue that a similar finite renormalization of the cosmological constant 
is necessary in pure gravity, and that this must be done when solving the 
effective field equations for 1-loop corrections to the graviton wave 
function and to the force of gravity.

\begin{flushleft}
\end{flushleft}

\vspace{1.5cm}

\begin{flushleft}
$^{*}$ e-mail: tsamis@physics.uoc.gr \\
$^{\dagger}$ e-mail: woodard@phys.ufl.edu \\
$^{\ddagger}$ e-mail: b.yesilyurt@ufl.edu
\end{flushleft}

\end{titlepage}

\section{Introduction}

One of the peculiarities of quantum gravity derives from the fact that the
inverse metric and the determinant of the metric are infinite order in the
graviton field. In consequence, counterterms generally affect all 1PI 
(one-particle-irreducible) n-point functions. For example, the $R^2$ 
counterterm which is needed to remove dimensionally regulated divergences
in the 1PI 2-graviton function \cite{tHooft:1974toh} also makes a finite, 
nonzero contribution to the 1-point function on de Sitter background to
recover the conformal anomaly \cite{Firouzjahi:2023wbe}.

The gravitational divergences induced by a single loop of massless matter 
fields can all be removed using two quadratic curvature counterterms,
\begin{equation}
\Delta \mathcal{L}_1 \equiv \alpha C_{\rho\sigma\mu\nu} C^{\rho\sigma\mu\nu}
\sqrt{-g} \qquad , \qquad \Delta \mathcal{L}_2 \equiv \beta R^2 \sqrt{-g} 
\; , \label{DL12}
\end{equation}
where $C_{\rho\sigma\mu\nu}$ is the Weyl tensor and $R$ is the Ricci scalar
\cite{Barvinsky:1985an}. For a massless, minimally coupled scalar the 
coefficients $\alpha$ and $\beta$ can be chosen as,
\begin{eqnarray}
\alpha &\!\!\! = \!\!\!& \frac{\mu^{D-4} \Gamma(\frac{D}2)}{2^8 \pi^{\frac{D}2}}
\frac{2}{(D\!+\!1) (D\!-\!1) (D\!-\!3)^2 (D\!-\!4)} \; , \label{alpha} \qquad \\
\beta &\!\!\! = \!\!\!& \frac{\mu^{D-4} \Gamma(\frac{D}2)}{2^8 \pi^{\frac{D}2}}
\frac{(D\!-\!2)}{(D\!-\!1)^2 (D\!-\!3) (D\!-\!4)} \; , \label{beta} \qquad
\end{eqnarray}
where $D$ is the dimension of spacetime and $\mu$ is the mass scale of 
dimensional regularization \cite{Park:2010pj,Park:2011ww}. On de Sitter 
background the Weyl ($C^2$) term makes no change in the 1-point function 
while the Eddington ($R^2$) term makes a finite change which, however, 
leaves a nonzero remainder. The purpose of this note is to show that 
preserving the Ward identity for the 1PI 2-point function requires that 
we make an additional, finite renormalization of the cosmological constant 
which nulls the 1-point function.

In section 2 we derive the connection between the 1-point function and the 
Ward identity for the primitive contribution to the 1PI 2-point function. 
Section 3 analyzes the contributions from the two required counterterms 
(\ref{DL12}), and from a cosmological counterterm. Our conlcusions are given
in section 4, including a discussion of the implications for pure gravity.

\section{Primitive Obstacle to Conservation}

The purpose of this section is to derive the connection between a nonzero
1-point function and the failure of the primitive 1PI 2-point function to
obey the Ward identity. We begin by defining the graviton field and giving
the scalar action and its variations. The obstacle is derived using a 
functional integral representation for the 1PI 2-point function which is
valid for an arbitrary homogeneous and isotropic background. The section
closes by evaluating the obstacle for de Sitter background.

\subsection{Preliminaries}

We define the graviton field $h_{\mu\nu}(x)$ by conformally rescaling with 
the time-dependent scale factor $a(x^0)$,
\begin{equation}
g_{\mu\nu} \equiv a^2 \widetilde{g}_{\mu\nu} \equiv a^2 \Bigl[\eta_{\mu\nu}
+ \kappa h_{\mu\nu}\Bigr] \qquad , \qquad \kappa^2 \equiv 16 \pi G \quad ,
\quad H \equiv \frac{\partial_0 a}{a^2} \; . 
\end{equation}
We do not at this stage take the de Sitter limit of constant Hubble parameter 
$H$. The massless, minimally coupled scalar action is definied in $D$ spacetime
dimensions to facilitate the use of dimensional regularization,
\begin{equation}
S[\phi,h] = -\frac12 \int \!\! d^Dx \, a^{D-2} \sqrt{-\widetilde{g}} \, 
\widetilde{g}^{\mu\nu} \partial_{\mu} \phi \partial_{\nu} \phi \; . 
\label{action}
\end{equation}
After taking its variations with respect to the scalar and the graviton we
set the graviton field to zero,
\begin{equation}
\frac{\delta S[\phi,0]}{\delta \phi(x)} = \partial_{\alpha} \Bigl[ a^{D-2} 
\partial^{\alpha} \phi\Bigr] \quad , \quad \frac{\delta S[\phi,0]}{\delta 
h_{\mu\nu}(x)} = \frac{\kappa}{2} a^{D-2} \Bigl[\partial^{\mu} \phi \partial^{\nu} 
\phi - \frac12 \eta^{\mu\nu} \partial^{\alpha} \phi \partial_{\alpha} \phi \Bigr] 
\; . \label{var1}
\end{equation}
Note that derivative indices are raised and lowered with the Minkowski metric, 
$\partial^{\mu} \equiv \eta^{\mu\nu} \partial_{\nu}$. The second variation 
with respect to the graviton is,
\begin{eqnarray}
\lefteqn{ \frac{\delta^2 S[\phi,0]}{\delta h_{\mu\nu}(x) \delta h_{\rho\sigma}(x')} 
= \frac{\kappa^2}{2} a^{D-2} \Biggl\{ -\frac14 \eta^{\alpha\beta} \eta^{\rho\sigma}
\eta^{\mu\nu} + \frac12 \eta^{\alpha\beta} \eta^{\mu (\rho} \eta^{\sigma)\nu} +
\frac12 \eta^{\alpha\rho} \eta^{\beta\sigma} \eta^{\mu\nu} } \nonumber \\ 
& & \hspace{3.4cm} + \frac12 \eta^{\alpha\mu} \eta^{\beta\nu} \eta^{\rho\sigma}
\!-\! 2 \eta^{\alpha (\rho} \eta^{\sigma) (\mu} \eta^{\nu) \beta} \! \Biggr\} 
\partial_{\alpha} \phi \partial_{\beta} \phi \delta^D(x \!-\! x') \; . \qquad 
\label{var2}
\end{eqnarray}

\subsection{Deriving the Obstacle}

It is most convenient to use a functional integral definition for the primitive
contribution to the 1PI graviton 2-point function,
\begin{equation}
-i\Bigl[\mbox{}^{\mu\nu} \Sigma^{\rho\sigma}_{\rm prim}\Bigr](x;x') = \Fint [d\phi]
e^{i S[\phi,0]} \Biggl\{ \frac{i \delta S[\phi,0]}{\delta h_{\mu\nu}(x)} 
\frac{i \delta S[\phi,0]}{\delta h_{\rho\sigma}(x')} + \frac{i \delta^2 S[\phi,0]}{
\delta h_{\mu\nu}(x) \delta h_{\rho\sigma}(x')} \Biggr\} . \label{Sigmaprim}
\end{equation}
Note that the integration is solely with respect to the scalar, even though the
integrand involves variations with respect to the graviton. The Ward identity for
a matter loop contribution to the graviton self-energy follows from stress-energy
conservation and takes the form,
\begin{equation}
\mathcal{W}^{\mu}_{~\alpha\beta} \!\times\! -i\Bigl[\mbox{}^{\alpha\beta} 
\Sigma^{\rho\sigma}_{\rm total}\Bigr](x;x') = 0 \; , \label{cons}
\end{equation} 
where we define the Ward operator as,
\begin{equation}
\mathcal{W}^{\mu}_{~\alpha\beta} \equiv \delta^{\mu}_{~(\alpha} \partial_{\beta)}
+ a H \delta^{\mu}_{~0} \eta_{\alpha\beta} \; . \label{Ward}
\end{equation}
Note that stress-energy conservation implies that the Ward operator annihilates
the graviton variation when the scalar obeys its equation of motion, 
\begin{equation}
\mathcal{W}^{\mu}_{~\alpha\beta} \!\times\! \frac{i \delta S[\phi,0]}{\delta 
h_{\alpha\beta}(x)} = \frac{\kappa}{2} \partial^{\mu} \phi(x) \!\times\!
\frac{i \delta S[\phi,0]}{\delta \phi(x)} \; . \label{step1}
\end{equation} 

The simplest and most general derivation of the obstacle is to perform a 
functional partial integration of the Ward operator acting on the first term,
\begin{eqnarray}
\lefteqn{ \frac{\kappa}{2} \partial^{\mu} \phi(x) \frac{i \delta S[\phi,0]}{
\delta \phi(x)} \frac{i \delta S[\phi,0]}{\delta h_{\rho\sigma}(x')} 
e^{i S[\phi,0]} = \frac{\kappa}{2} \partial^{\mu} \phi(x) \frac{\delta}{
\delta \phi(x)} \Biggl[ e^{i S[\phi,0]}\Biggr] \frac{i \delta S[\phi,0]}{\delta 
h_{\rho\sigma}(x')} } \nonumber \\
& & \hspace{5cm} \longrightarrow -\frac{\kappa}{2} e^{i S[\phi,0]} 
\frac{\delta}{\delta \phi(x)} \Biggl[ \partial^{\mu} \phi(x) \frac{i \delta 
S[\phi,0]}{\delta h_{\rho\sigma}(x')} \Biggr] \; . \qquad \label{partint}
\end{eqnarray}
There is no contribution from the variation of the $\partial^{\mu} \phi(x)$
term in dimensional regularization but the variation of the second term gives,
\begin{eqnarray}
\lefteqn{-\frac{\kappa}{2} \partial^{\mu} \phi(x) \!\times\! 
\frac{i \delta^2 S[\phi,0]}{\delta \phi(x) \delta h_{\rho\sigma}(x')} } 
\nonumber \\
& & \hspace{0cm} = \frac{i \kappa^2}{2} \partial^{\mu} \phi \Biggl\{ 
\partial^{(\rho} \! \Bigl[ a^{D-2} \partial^{\sigma)} \phi \delta^D(x \!-\! x') 
\Bigr] \!-\! \frac12 \eta^{\rho\sigma} \partial_{\alpha} \Bigl[ a^{D-2} 
\partial^{\alpha} \phi \delta^D(x \!-\! x')\Bigr] \! \Biggr\} . \qquad 
\label{W3}
\end{eqnarray}
Acting the Ward operator on the second part of (\ref{Sigmaprim}) gives,
\begin{eqnarray}
\lefteqn{ \mathcal{W}^{\mu}_{~\alpha\beta} \!\times\! \frac{i \delta^2 
S[\phi,0]}{\delta h_{\alpha\beta}(x) \delta h_{\rho\sigma}(x')} = 
\frac{i \kappa^2}{2} \Biggl\{ -\partial_{\alpha} \Bigl[ a^{D-2} 
\partial^{\alpha} \phi \eta^{\mu (\rho} \partial^{\sigma)} \phi 
\delta^D(x \!- \! x') \Bigr] } \nonumber \\
& & \hspace{0cm} + \frac12 \eta^{\mu (\rho} \partial^{\sigma)} \Bigl[
a^{D-2} \partial_{\alpha} \phi \partial^{\alpha} \phi \delta^D(x \!-\! x')\Bigr]
- \partial^{(\rho} \Bigl[ a^{D-2} \partial^{\sigma)} \phi \partial^{\mu} \phi
\delta^D(x \!- x') \Bigr] \nonumber \\
& & \hspace{0cm} + \frac12 \eta^{\rho\sigma} \partial_{\alpha} \Bigl[a^{D-2} 
\partial^{\alpha} \phi \partial^{\mu} \phi \delta^D(x \!-\! x') \Bigr]
+ \frac12 a^{D-4} \partial^{\mu} \Bigl[ a^{2} \partial^{\rho} 
\phi \partial^{\sigma} \phi \delta^D(x \!-\! x') \Bigr]  \nonumber \\
& & \hspace{5.3cm} -\frac14 a^{D-4} \eta^{\rho\sigma} \partial^{\mu} \Bigl[a^{2}
\partial_{\alpha} \phi \partial^{\alpha} \phi \delta^D(x \!-\! x')\Bigr]
\Biggr\} . \qquad \label{W4}
\end{eqnarray}
After some judicious manipulations, the sum of (\ref{W3}) and (\ref{W4}) becomes,
\begin{eqnarray} 
\lefteqn{ \mathcal{W}^{\mu}_{~\alpha\beta} \!\times\! -i \Bigl[\mbox{}^{\alpha\beta}
\Sigma^{\rho\sigma}_{\rm prim}\Bigr](x;x') = - \kappa \Fint [d\phi] e^{iS[\phi,0]} } 
\nonumber \\
& & \hspace{0cm} \times \Biggl\{ \! \eta^{\mu (\rho} \delta^{\sigma)}_{~~\alpha}
\partial_{\beta} \Bigl[ \frac{i \delta S[\phi,0]}{\delta h_{\alpha\beta}(x)}
\delta^D(x \!-\! x') \Bigr] - \frac1{2 a^2} \frac{i \delta S[\phi,0]}{\delta 
h_{\rho\sigma}(x)} \partial^{\mu} \Bigl[ a^2 \delta^D(x \!-\! x')\Bigr] \! 
\Biggr\} . \qquad \label{obstacleprim}
\end{eqnarray}
Note that this expression does not require de Sitter; it is valid for an arbitrary
cosmological background.

\subsection{Evaluating the Obstacle on de Sitter}

The obstacle (\ref{obstacleprim}) can be expressed in terms of the scalar 
propagator $i\Delta(x;x')$,
\begin{eqnarray}
\lefteqn{ \mathcal{W}^{\mu}_{~\alpha\beta} \!\times\! -i \Bigl[\mbox{}^{\alpha\beta}
\Sigma^{\rho\sigma}_{\rm prim}\Bigr](x;x') } \nonumber \\
& & \hspace{0cm} = \frac{i \kappa^2}{2} \Biggl\{-\eta^{\mu (\rho} \delta^{\sigma)}_{
~~\alpha} \partial_{\beta} \Bigl[ a^{D-2} \Bigl( \eta^{\alpha\gamma} \eta^{\beta\delta} 
- \frac12 \eta^{\alpha\beta} \eta^{\gamma\delta} \Bigr) \partial_{\gamma} 
\partial'_{\delta} i\Delta(x;x') \delta^D(x \!-\! x') \Bigr] \nonumber \\
& & \hspace{.7cm} + \frac{a^{D-4}}{2} \Bigl( \eta^{\rho\gamma} \eta^{\sigma\delta}
- \frac12 \eta^{\rho\sigma} \eta^{\gamma\delta} \Bigr) \partial_{\gamma} 
{\partial'}_{\delta} i\Delta(x;x')\Bigl\vert_{x'=x} \partial^{\mu} \Bigl[
a^2 \delta^D(x \!-\! x') \Bigr] \! \Biggr\} . \qquad \label{obstprop}
\end{eqnarray}
This expression is still valid for a general cosmological background.

We now specialize to de Sitter for which the Hubble parameter is constant.
The scalar propagator is quite complicated in $D$ dimensions, however, the 
coincidence limit of its mixed second derivative is simple \cite{Onemli:2002hr,
Onemli:2004mb},
\begin{equation}
\partial_{\mu} \partial'_{\nu} i\Delta(x;x') \Bigl\vert_{x' = x} = -\Bigl(
\frac{D\!-\!1}{D} \Bigr) k H^2 a^2 \eta_{\mu\nu} \;\; , \;\; k \equiv
\frac{H^{D-2}}{(4\pi)^{\frac{D}2}} \frac{\Gamma(D \!-\! 1)}{\Gamma(\frac{D}2)}
\; . \label{prop}
\end{equation}
Substituting (\ref{prop}) in (\ref{obstprop}) gives the obstacle for de
Sitter,
\begin{eqnarray}
\lefteqn{ \mathcal{W}^{\mu}_{~\alpha\beta} \!\times\! -i \Bigl[
\mbox{}^{\alpha\beta} \Sigma^{\rho\sigma}_{\rm prim}\Bigr](x;x') 
\longrightarrow i \kappa^2 H^2 k \frac{(D\!-\!2) (D\!-\!1)}{4 D} } 
\nonumber \\
& & \hspace{1.7cm} \times \Biggl\{ -\eta^{\mu (\rho} \partial^{\sigma)} 
\Bigl[ a^D \delta^D(x \!-\! x')\Bigr] + \frac12 a^{D-2} \eta^{\rho\sigma} 
\partial^{\mu} \Bigl[ a^2 \delta^D(x\!-\!x') \Bigr] \Biggr\} . \qquad  
\label{obstdS}
\end{eqnarray}
We have retained dimensional regularization even though the result is finite.

\section{Counterterms}

The purpose of this section is to consider how counterterms affect the obstacle.
We begin by demonstrating that the Weyl counterterm does not affect the obstacle 
at all, while the Eddington counterterm does not completely eliminate it. The
section closes by showing that a finite cosmological counterterm absorbs the
obstacle. 

\subsection{Weyl}

The Weyl counterterm has the most complicated tensor structure but it makes the
simplest contribution to the graviton self-energy. We define a second order
tensor differential operator $\mathcal{C}_{\alpha\beta\gamma\delta}^{~~~~~\mu\nu}$
by expanding the Weyl tensor of the conformally rescaled metric $\widetilde{g}_{
\mu\nu}$ in powers of the graviton field,
\begin{equation}
\widetilde{C}_{\alpha\beta\gamma\delta} = \mathcal{C}_{\alpha\beta\gamma\delta}^{
~~~~~\mu\nu} \!\times\! \kappa h_{\mu\nu} + O(\kappa^2 h^2) \; . \label{Cdef}
\end{equation}
The explicit form of this operator can be found in section 3.2 of 
\cite{Leonard:2014zua} but we require only the fact that it is both transverse 
and traceless. The first two variations of the Weyl counterterm are,
\begin{equation}
\frac{i \delta \Delta S_1[0]}{\delta h_{\mu\nu}(x)} = 0 \;\; , \;\; 
\frac{i \delta^2 \Delta S_1[0]}{\delta h_{\mu\nu}(x) \delta h_{\rho\sigma}(x')}
= 2 i \alpha \kappa^2 \mathcal{C}^{\alpha\beta\gamma\delta \mu\nu} \Bigl[ a^{D-4}
\mathcal{C}_{\alpha\beta\gamma\delta}^{~~~~~\rho\sigma} \delta^D(x \!-\! x')
\Bigr] \; . \label{varS1}
\end{equation}
It follows that the Ward operator annihilates the contribution from the Weyl
counterterm,
\begin{equation}
\mathcal{W}^{\mu}_{~\alpha\beta} \!\times\! -i \Bigl[\mbox{}^{\alpha\beta}
\Sigma^{\rho\sigma}_{1}\Bigr](x;x') = 0 \; .
\end{equation}
Note that all of these relations pertain for any cosmological background.

\subsection{Eddington}

It is best specialize the Eddington counterterm to de Sitter background (with
constant $H$) and to break it up into three terms involving the cosmological 
constant $\Lambda \equiv (D-1) H^2$,
\begin{equation}
R^2 = \Bigl[R - D \Lambda\Bigr]^2 + 2 D \Lambda \Bigl[R - (D\!-\!2) \Lambda
\Bigr] + D (D \!-\! 4) \Lambda^2 \; . \label{Rsq123}
\end{equation}
This leads to three counterterms,
\begin{eqnarray}
\Delta \mathcal{L}_{2a} &\!\!\! = \!\!\!& \beta \Bigl[R - D \Lambda\Bigr]^2
\sqrt{-g} \; , \qquad \label{DL2a} \\
\Delta \mathcal{L}_{2b} &\!\!\! = \!\!\!& 2 D \beta \Lambda \Bigl[ R -
(D\!-\!2) \Lambda\Bigr] \sqrt{-g} \; , \qquad \label{DL2b} \\
\Delta \mathcal{L}_{2c} &\!\!\! = \!\!\!& D (D\!-\!4) \beta \Lambda^2 \sqrt{-g}
\; . \qquad \label{DL2c}
\end{eqnarray}
Because $\Delta \mathcal{L}_{2c}$ is the same as a cosmological constant, we
will postpone its consideration until the next subsection.

To analyze $\Delta \mathcal{L}_{2a}$ we expand $R - D \Lambda$ in powers of 
the graviton field, like what we did with the conformally rescaled Weyl tensor 
in expression (\ref{Cdef}),
\begin{equation}
R - D \Lambda = \frac1{a^2} \overline{\mathcal{F}}^{\mu\nu} \!\times\!
\kappa h_{\mu\nu} + O(\kappa^2 h^2) \; . \label{Rexp}
\end{equation}
The tensor differential operator $\overline{\mathcal{F}}^{\mu\nu}$ is,
\begin{eqnarray}
\lefteqn{ \overline{\mathcal{F}}^{\mu\nu} \equiv \partial^{\mu} \partial^{\nu}
- \eta^{\mu\nu} \Bigl[\partial^2 - (D\!-\!1) a H \partial_0\Bigr] } 
\nonumber \\
& & \hspace{4cm} - 2 (D\!-\!1) a H \delta^{(\mu}_{~~0} \partial^{\nu)} + 
D (D\!-\!1) a^2 H^2 \delta^{\mu}_{~0} \delta^{\nu}_{~0} \; . \qquad 
\label{Fbar}
\end{eqnarray}
Variation of $\Delta S_{2a}$ results in this operator being partially 
integrated to give,
\begin{eqnarray}
\lefteqn{ \mathcal{F}^{\mu\nu} = \partial^{\mu} \partial^{\nu} - \eta^{\mu\nu}
\Bigl[\partial^2 + (D\!-\!1) a H \partial_0 + (D\!-\!1) a^2 H^2 \Bigr] } 
\nonumber \\
& & \hspace{3cm} + 2 (D\!-\!1) a H \delta^{(\mu}_{~~0} \partial^{\nu)} + 
(D\!-\!2) (D\!-\!1) a^2 H^2 \delta^{\mu}_{~0} \delta^{\nu}_{~0} \; . \qquad 
\label{Fdef}
\end{eqnarray}
Because $\Delta \mathcal{L}_{2a}$ is quadratic in the graviton, its first
variation vanishes at $h_{\mu\nu} = 0$. Its second variation is,
\begin{equation}
\frac{i \delta^2 \Delta S_{2a}[0]}{\delta h_{\mu\nu}(x) \delta 
h_{\rho\sigma}(x')} = 2 i \kappa^2 \beta \mathcal{F}^{\mu\nu} \Bigl[a^{D-4} 
\mathcal{F}^{\rho\sigma} \delta^D(x \!-\! x')\Bigr] \; . \label{varS2a}
\end{equation}
The Ward operator annihilates this contribution,
\begin{equation}
\mathcal{W}^{\mu}_{~\alpha\beta} \!\times\! -i \Bigl[\mbox{}^{\alpha\beta}
\Sigma^{\rho\sigma}_{2a}\Bigr](x;x') = 0 \; .
\end{equation}

The middle part of the Eddington counterterm $\Delta \mathcal{L}_{2b}$ is
proportional to the Einstein-Hilbert Lagrangian so its first variation 
vanishes at $h_{\mu\nu} = 0$. Its second variation is,
\begin{eqnarray}
\lefteqn{ \frac{i \delta^2 \Delta S_{2b}[0]}{\delta h_{\mu\nu}(x) \delta 
h_{\rho\sigma}(x')} = D (D\!-\!1) i \kappa^2 H^2 \beta \Biggl\{ \!
\Bigl[ \eta^{\mu (\rho} \eta^{\sigma) \nu} \!\!-\! \eta^{\mu\nu} 
\eta^{\rho\sigma}\Bigr] \partial^{\alpha} \! \Bigl[ a^{D-2} \partial_{\alpha}
\delta^D(x \!-\! x') \Bigr] } \nonumber \\
& & \hspace{2cm} + \Bigl[ 2 {\partial'}^{(\mu} \eta^{\nu)(\rho} 
\partial^{\sigma)} \!+\! \eta^{\mu\nu} \partial^{\rho} \partial^{\sigma} \!+\! 
{\partial'}^{\mu} {\partial'}^{\nu} \eta^{\rho\sigma}\Bigr] \Bigl[ a^{D-2}
\delta^D(x \!-\! x')\Bigr] \! \Biggr\} . \qquad \label{varS2b}
\end{eqnarray}
\newpage
\noindent The Ward operator also annihilates this contribution,
\begin{equation}
\mathcal{W}^{\mu}_{~\alpha\beta} \!\times\! -i \Bigl[\mbox{}^{\alpha\beta}
\Sigma^{\rho\sigma}_{2b}\Bigr](x;x') = 0 \; .
\end{equation}

\subsection{Cosmological}

The cosmological counterterm is,
\begin{equation}
\Delta \mathcal{L}_3 \equiv \gamma \sqrt{-g} = \gamma a^D 
\sqrt{-\widetilde{g}} \; . \label{DL3}
\end{equation}
This term can obviously be combined with $\Delta \mathcal{L}_{2c}$, leading 
to a counterterm of the same form as (\ref{DL3}) but with coefficient,
\begin{equation}
\gamma' = \gamma + D (D\!-\!1)^2 (D\!-\!4) H^4 \beta \; . \label{gammap}
\end{equation}
The first variation is,
\begin{equation}
\frac{i \delta \Delta S_{2c + 3}[0]}{\delta h_{\mu\nu}} = 
\frac{i \gamma' \kappa}{2} a^D \eta^{\mu\nu} \qquad \Longrightarrow \qquad 
\mathcal{W}^{\mu}_{~\alpha\beta} \!\times\! \frac{i \delta \Delta S_{2c + 3}[0]}{
\delta h_{\alpha\beta}(x)} = 0 \; .
\end{equation}
The second variation is,
\begin{equation}
\frac{i \delta^2 \Delta S_{2c + 3}[0]}{\delta h_{\mu\nu}(x) \delta 
h_{\rho\sigma}(x')} = \frac{i \gamma' \kappa^2}{2} a^D \Bigl[- \eta^{\mu (\rho} 
\eta^{\sigma) \nu} + \frac12 \eta^{\mu\nu} \eta^{\rho\sigma}\Bigr] 
\delta^D(x \!-\! x') \; . 
\end{equation}
Acting the Ward operator on this gives, 
\begin{eqnarray}
\lefteqn{\mathcal{W}^{\mu}_{~\alpha\beta} \!\times\! \frac{i \delta^2 
\Delta S_{2c + 3}[0]}{\delta h_{\alpha\beta}(x) \delta h_{\rho\sigma}(x')} } 
\nonumber \\
& & \hspace{0.9cm} = \frac{i \gamma' \kappa^2}{2} \Biggl\{\! -\eta^{\mu (\rho} 
\partial^{\sigma)} \Bigl[ a^D \delta^D(x \!-\! x')\Bigr] \!+\! \frac12 a^{D-2} 
\eta^{\rho\sigma} \partial^{\mu} \Bigl[ a^2 \delta^D(x\!-\!x') \Bigr] 
\! \Biggr\} . \qquad \label{obstcosmo}
\end{eqnarray}

Because the obstruction from a cosmological constant (\ref{obstcosmo}) takes
the same form as the primitive obstruction (\ref{obstdS}), they can be made 
to cancel by choosing,
\begin{equation}
\gamma' = -\frac{(D\!-\!2) (D\!-\!1)}{2D} k H^2 \; , \label{gammaprime}
\end{equation}
where the constant $k$ was defined in expression (\ref{prop}). From 
(\ref{gammaprime}) we see that this corresponds to a renormalization
of the cosmological constant which is finite in $D=4$ dimensions,
\begin{equation}
\gamma = -\frac{\mu^{D-4} H^4}{2^7 \pi^{\frac{D}2}} \frac{(D\!-\!2) 
\Gamma(\frac{D}2 \!+\! 1)}{D \!-\!3} - \frac{H^D}{(4\pi)^{\frac{D}2}}
\frac{(D\!-\!2) \Gamma(D)}{4 \Gamma(\frac{D}2 \!+\! 1)} \longrightarrow
-\frac{H^4}{8 \pi^2} \; . \label{gamma}
\end{equation}

\section{Conclusions}

We considered contributions to the graviton self-energy $-i [\mbox{}^{\mu\nu}
\Sigma^{\rho\sigma}](x;x')$ from a loop of massless, minimally coupled scalars, 
at first on a general cosmological background and then specialized to de Sitter. 
Conservation of the scalar stress tensor is usually thought to imply that this 
quantity should obey a Ward identity (\ref{cons}). However, the primitive 
contribution to $-i [\mbox{}^{\mu\nu} \Sigma^{\rho\sigma}](x;x')$ suffers a 
delta function obstruction (\ref{obstacleprim}), whose specialization to de
Sitter is (\ref{obstdS}). This obstruction is proportional to the graviton 
1-point function on a general cosmological background, and is finite on de
Sitter.

Of course the primitive contribution contains divergences which must be 
subtracted using counterterms (\ref{DL12}) proportional to the squares of the
Weyl tensor and the Ricci scalar, with coefficients (\ref{alpha}-\ref{beta}).
The Weyl ($C^2$) counterterm has no effect on the obstruction. The 
Eddington ($R^2$) counterterm alters the obstruction but does not cancel 
it. Owing to global scale invariance in $D=4$ dimensions, the Eddington 
contribution to the obstruction, and to the graviton 1-point function, is 
finite on de Sitter. This means that the finite part of the coefficient 
(\ref{beta}) has no effect on either the obstruction or the 1-point function; 
only the divergent part matters. Full cancellation of the obstruction requires 
an additional renormalization of the cosmological constant (\ref{DL3}) with 
finite (on de Sitter) coefficient (\ref{gamma}). Making this renormalization 
also causes the graviton 1-point function to vanish.

Of course we must subtract divergences but finite renormalizations are 
usually considered to be optional, so we should explain the strong motivation 
for making this one. First, is the fact that it is needed to remove the 
obstruction to the Ward identity (\ref{cons}). This is not a sterile,
mathematical problem, it compromises our ability to use the renormalized 
graviton self-energy to solve for quantum corrections to gravitational 
radiation and to the force of gravity using the linearized effective field
equations \cite{Park:2015kua},
\begin{equation}
\mathcal{D}^{\mu\nu\rho\sigma} \!\times\! \kappa h_{\mu\nu}(x) -
\int \!\! d^4x' \Bigl[\mbox{}^{\mu\nu} \Sigma^{\rho\sigma}_{\rm ren}
\Bigr](x;x') \kappa h_{\rho\sigma}(x') = 8 \pi G T^{\mu\nu}(x) \; . 
\label{QEinstein}
\end{equation}
Here $T^{\mu\nu}(x)$ is the stress tensor density and $\mathcal{D}^{\mu\nu
\rho\sigma}$ is the Lichnerowitz operator. We need conservation because the 
Ward operator (\ref{Ward}) annihilates both the stress tensor density and 
the Lichnerowitz operator for de Sitter,
\begin{eqnarray}
\lefteqn{\mathcal{D}^{\mu\nu\rho\sigma} = \frac{a^2}{2} \Bigl[ (\eta^{\mu (\rho}
\eta^{\sigma) \nu} \!\!-\! \eta^{\mu\nu} \eta^{\rho\sigma}) \partial^2 \!+\!
\eta^{\mu\nu} \partial^{\rho} \partial^{\sigma} \!+\! \eta^{\rho\sigma}
\partial^{\mu} \partial^{\nu} \!-\! 2 \partial^{(\mu} \eta^{\nu) (\rho}
\partial^{\sigma)} \Bigr] + H a^3 } \nonumber \\
& & \hspace{-0.5cm} \times\! \Bigl[ (\eta^{\mu\nu} \eta^{\rho\sigma} \!\!-\!
\eta^{\mu (\rho} \eta^{\sigma) \nu} ) \partial_0 \!-\! 2 \eta^{\mu\nu}
\delta^{(\rho}_{~~0} \partial^{\sigma)} \!+\! 2 \delta^{(\rho}_{~~0} 
\eta^{\sigma) (\mu} \partial^{\nu)} \Bigr] \!+\! 3 H^2 a^4 \eta^{\mu\nu} 
\delta^{\rho}_{~0} \delta^{\sigma}_{~0} \; . \qquad \label{Lich}
\end{eqnarray}
Hence the equation (\ref{QEinstein}) would not be consistent if the Ward 
operator did not also annihilate $-i [\mbox{}^{\mu\nu} \Sigma^{\rho\sigma}_{
\rm ren}](x;x')$.

An additional motivation comes from the connection (\ref{obstacleprim})
between the obstacle and the graviton 1-point function. If the 1-point 
function fails to vanish then back-reaction will change the true expansion
rate so that it no longer coincides with the quantity we call the ``Hubble 
parameter'' \cite{Tsamis:2005je}. This does not represent a fine tuning 
but is instead similar to the on-shell renormalization condition that the 
quantity ``$m$'' in the electron propagator of quantum electrodynamics 
should represent the actual electron mass.

Relation (\ref{obstacleprim}) applies as well to a massive, minimally 
coupled scalar. However, the greatest relevance of this work is to the 
study of how loops of inflationary gravitons affect the graviton mode 
function \cite{Tan:2021lza} and the force of gravity \cite{Tan:2022xpn}. 
When graviton loops are involved one must act the Ward operator 
(\ref{Ward}) on each of the two points in order to reach zero 
\cite{Tan:2021ibs},
\begin{equation}
\mathcal{W}^{\mu}_{~\alpha\beta} \!\times\! \mathcal{W}^{\prime \rho
}_{~~\gamma\delta} \!\times\! -i \Bigl[\mbox{}^{\alpha\beta} \Sigma^{
\gamma\delta}_{\rm ren}\Bigr](x;x') = 0 \; .
\end{equation}
An explicit computation shows that this relation is obeyed for $x^{\mu} 
\neq {x'}^{\mu}$ \cite{Tsamis:1996qk}. However, it seems inevitable that
there will be a delta function obstacle, similar to the scalar relation 
(\ref{obstacleprim}), and that removing this obstacle will require a 
finite renormalization to cancel graviton loop contributions to the 
1-point function.

\vskip .5cm 

\centerline{\bf Acknowledgements}

This work was partially supported by NSF grant PHY-2207514 and by the 
Institute for Fundamental Theory at the University of Florida.

\end{document}